\documentclass[twocolumn]{aastex63}
\usepackage{threeparttable}
\usepackage{amssymb,amsmath}
\usepackage{graphicx}
\usepackage{CJKutf8}
\usepackage{natbib}
\usepackage{textcomp}
\usepackage{url}

\newcommand{\dn}{$\mbox{D}_n\mbox{4000}$}
\newcommand{\ewhd}{$\mbox{EW(H}\delta\mbox{)}$}

\begin{document}
\title{Towards precise galaxy evolution: a comparison between spectral indices of $z\sim1$ galaxies in the IllustrisTNG simulation and the LEGA-C survey }

\shorttitle{TNG v.s. LEGA-C}
\shortauthors{Wu et al.}

\author{Po-Feng Wu \begin{CJK*}{UTF8}{bkai}(吳柏鋒)\end{CJK*}}
\altaffiliation{East Asia Core Observatory Association Fellow}
\affiliation{National Astronomical Observatory of Japan, Osawa 2-21-1, Mitaka, Tokyo 181-8588, Japan}
\affiliation{Max-Planck-Institut f\"{u}r Astronomie, K\"{o}nigstuhl 17, D-69117, Heidelberg, Germany}
\affiliation{Institute of Astronomy \& Astrophysics, Academia Sinica, Taipei 10617, Taiwan}

\author{Dylan Nelson}
\affiliation{Universit\"{a}t Heidelberg, Zentrum f\"{u}r Astronomie, Institut f\"{u}r theoretische Astrophysik, Albert-Ueberle-Str. 2, 69120 Heidelberg, Germany}

\author{Arjen van der Wel}
\affiliation{Sterrenkundig Observatorium, Universiteit Gent, Krijgslaan 281 S9, B-9000 Gent, Belgium}
\affiliation{Max-Planck-Institut f\"{u}r Astronomie, K\"{o}nigstuhl 17, D-69117, Heidelberg, Germany}

\author{Annalisa Pillepich}
\affiliation{Max-Planck-Institut f\"{u}r Astronomie, K\"{o}nigstuhl 17, D-69117, Heidelberg, Germany}

\author{Stefano Zibetti}
\affiliation{INAF-Osservatorio Astrofisico di Arcetri, Largo Enrico Fermi 5, I-50125 Firenze, Italy}

\author{Rachel Bezanson}
\affiliation{University of Pittsburgh, Department of Physics and Astronomy, 100 Allen Hall, 3941 O'Hara St, Pittsburgh PA 15260, USA}

\author{Francesco D’Eugenio}
\affiliation{Sterrenkundig Observatorium, Universiteit Gent, Krijgslaan 281 S9, B-9000 Gent, Belgium}

\author{Anna Gallazzi}
\affiliation{INAF-Osservatorio Astrofisico di Arcetri, Largo Enrico Fermi 5, I-50125 Firenze, Italy}

\author{Camilla Pacifici}
\affiliation{Space Telescope Science Institute, 3700 San Martin Drive, Baltimore, MD 21218, USA}

\author{Caroline M. S. Straatman}
\affiliation{Sterrenkundig Observatorium, Universiteit Gent, Krijgslaan 281 S9, B-9000 Gent, Belgium}

\author{Ivana Bari\v{s}i\'{c}}
\affiliation{Max-Planck-Institut f\"{u}r Astronomie, K\"{o}nigstuhl 17, D-69117, Heidelberg, Germany}

\author{Eric F. Bell}
\affiliation{Department of Astronomy, University of Michigan, 1085 South University Avenue, Ann Arbor, MI 48109-1107, USA}

\author{Michael V. Maseda}
\affiliation{Leiden Observatory, Leiden University, P.O. Box 9513, 2300 RA, Leiden, The Netherlands}

\author{Adam Muzzin}
\affiliation{Department of Physics and Astronomy, York University, 4700 Keele St., Toronto, Ontario, M3J 1P3, Canada}

\author{David Sobral}
\affiliation{Physics Department, Lancaster University, Lancaster LA1 4YB, UK}
\affiliation{Leiden Observatory, Leiden University, PO Box 9513, 2300 RA Leiden, The Netherlands}

\author{Katherine E. Whitaker}
\affiliation{Department of Astronomy, University of Massachusetts, Amherst, MA 01003, USA}
\affiliation{Cosmic Dawn Center (DAWN)}

\correspondingauthor{Po-Feng Wu}
\email{pfwu@asiaa.sinica.edu.tw}

\begin{abstract}
    We present the first comparison of observed stellar continuum spectra of high-redshift galaxies and mock galaxy spectra generated from hydrodynamical simulations. The mock spectra are produced from the IllustrisTNG TNG100 simulation combined with stellar population models and take into account dust attenuation and realistic observational effects (aperture effects and noise). We compare the simulated \dn\ and \ewhd\ of galaxies with $10.5 \leq \log(M_\ast/M_\odot) \leq 11.5$ at $0.6 \leq z \leq 1.0$ to the observed distributions from the LEGA-C survey. TNG100 globally reproduces the observed distributions of spectral indices, implying that the age distribution of galaxies in TNG100 is generally realistic. Yet there are small but significant differences. For old galaxies, TNG100 shows small \dn\ when compared to LEGA-C, while LEGA-C galaxies have larger \ewhd\ at fixed \dn. There are several possible explanations: 1) LEGA-C galaxies have overall older ages combined with small contributions (a few percent in mass) from younger ($<1$~Gyr) stars, while TNG100 galaxies may not have such young sub-populations; 2) the spectral mismatch could be due to systematic uncertainties in the stellar population models used to convert stellar ages and metallicities to observables. In conclusion, the latest cosmological galaxy formation simulations broadly reproduce the global age distribution of galaxies at $z\sim1$ and, at the same time, the high quality of the latest observed and simulated datasets help constrain stellar population synthesis models as well as the physical models underlying the simulations. 
\end{abstract}

\keywords{}

\section{Introduction}

Under the current $\Lambda$ Cold Dark Matter ($\Lambda$CDM) cosmological paradigm, galaxies form out of the density field in the early Universe. Gravity amplifies the initially weak fluctuations and shapes the large-scale structure with galaxies forming at the nodes of the cosmic web \citep{whi78,blu84}. 
At the same time, baryonic physics, e.g., the formation of dense molecular clouds, conversion from gas to stars, growth of supermassive black holes, and their injection of energy and momentum to the ambient environment, vigorously operate on smaller scales at the formation sites of galaxies \citep{kat91,kat92a,kat92b,spr05b,som15}. The interplay between the gravitational field of the large-scale structure and the baryonic physics on smaller scales governs the formation and evolution of galaxies over cosmic time. 

To decipher the galaxy formation process, cosmological hydrodynamical simulations start from initial conditions near the beginning of the Universe, incorporate a large number of physical mechanisms based on effective models, and collectively produce a galaxy population that is in reasonable agreement with a selected set of observational constraints \citep[e.g.,][]{som15,sch15,pil18}. Properties that are not used for calibrating the input parameters are thus predictive in nature and can either further validate the simulation or guide observational work. 
Recent simulation projects have had great success and demonstrated that large volume hydrodynamical simulations with kilo-parsec resolution are able to reproduce not only the observational constraints that they are tuned for, but also several other scaling relations and fundamental properties of observed galaxies \citep[e.g.,][]{dub14,vog14,sch15,kha15,dave16}.

One of the important achievements is recovering the color distribution of local galaxies, which is a fundamental observable and a proxy to access the population of constituent stars of galaxies. Both the EAGLE \citep{sch15} and the IllustrisTNG simulations \citep{pil18} produce a bimodal $g-r$ color distribution and approximately correct fractions of red and blue galaxies \citep{tra15,tra17,nel18}. Going beyond colors, \citet{tra17} showed a similarity between the D4000 distributions of observed and EAGLE galaxies. 

These successes in matching the colors and spectral absorption features of the local galaxy population give us confidence that the constituent stellar populations in simulated galaxies resemble the observed galaxies. The formation histories of simulated galaxies inform us about the formation of galaxies in the real Universe.  
Meanwhile, the level of agreement between observed and simulated populations among the red sequence, blue cloud, and green valley can indicate ways to improve the physical models that drive the formation histories of galaxies in simulations \citep{tra15,tra17,nel18,ang21}.

Furthermore, the aforementioned validation has primarily been undertaken for local galaxies. Even if the simulated populations resemble the observed ones at $z\sim0$, they may have taken different evolutionary pathways. Of particular interest are whether simulated galaxy populations beyond the local Universe contain enough massive quiescent galaxies \citep{don19, don21b} or lack the reddest galaxies \citep{aki21}. Meanwhile, some apparent inconsistencies with models, e.g., in the specific SFRs \citep{kav17,don19} of galaxies at $z\gtrsim1$, appear to be resolved by the careful stellar population modeling of the observational data \citep{nel21}.

Comparing the constituent stellar populations of simulated galaxies to observations of galaxies in the earlier Universe is an essential milestone towards validating and further tuning the physical models of cosmological hydrodynamical simulations. As a step towards the goal, we aim to access the stellar population of galaxies in the IllustrisTNG simulation at a large look-back time through the two most widely used spectral features, \dn\ \citep{bal99} and \ewhd\ \citep[H$\delta_a$ in][]{wor97}. The recently finished Large Early Galaxy Astrophysics Census (LEGA-C) \citep{vdw16} provides observations of both high signal-to-noise (S/N) spectra and enough statistical power to accurately quantify the distributions of spectra features of ancient galaxy populations. The survey targets $\sim3000$ massive galaxies at $0.6 < z < 1.0$, corresponding to a look-back time of $6-8$~Gyrs, when the Universe was half of its current age. The quality and the quantity of the spectra obtained by the LEGA-C survey puts an extra anchor point on the stellar populations of galaxies at large look-back times. 

In Section~\ref{sec:data} we describe the simulated and the observation data, and the construction of the mock spectra to link the two together. Section~\ref{sec:result} presents the comparison to observations and Section~\ref{sec:dis} discusses the implications. We then summarize the paper in Section~\ref{sec:sum}. 

\section{Data and Analysis}
\label{sec:data}
\subsection{Observations: LEGA-C survey}

The LEGA-C survey uses the Visible Multi-Object Spectrograph \citep[VIMOS,][]{lef03} mounted on the 8--m Very Large Telescope to obtain rest-frame optical spectra of $\sim4000$ $K_s$-band-selected galaxies, mainly with $M_\ast \gtrsim 10^{10} M_\odot$ at $0.6 < z < 1.0$ \citep{vdw16}. Each galaxy receives $\sim20$ hr of on-source exposure time at a spectral resolution of $R\sim3500$, achieving a typical continuum $S/N$ of 20\AA$^{-1}$ \citep{str18}.

From the final LEGA-C data release (van der Wel et al. submitted), we start from the 2915 galaxies with $0.6 < z < 1.0$ that are LEGA-C primary targets, which have estimated volume completeness correction factors \citep{str18}. We then select 2319 galaxies with stellar mass $10.5 \leq \log(M_\ast/M_\odot) \leq 11.5$. The lower mass limit ensures that the magnitude limit of the survey does not introduce a strong bias against low-mass red galaxies. We also require that the spectra cover the wavelength range for measuring the \dn\ and \ewhd\ features and are thus left with 1963 galaxies.  
We finally exclude 76 galaxies whose photometry is contaminated by near neighbors, making the inferred stellar masses unreliable. This is $<4\%$ of the sample and we expect a negligible effect on the final statistics. In total, 1887 galaxies fulfill the selection criteria. 

We derive galaxy stellar masses by fitting the observed spectral energy distributions (SEDs) from the UltraVISTA catalog \citep{muz13a}. We take broadband photometry from B, V, r', i', z', Y, and J filters. Based on our own recalibration (van der Wel et al. submitted), we add 0.12, 0.05, 0.04, 0.04, 0.05, 0.07, and 0.07 magnitudes to each band, respectively. The new zeropoints are determined by matching the colors of stars in the COSMOS field, contrary to the comparison with the \citet{bc03} model templates used by \citet{muz13a}. 

We generate the SED templates using the Flexible Stellar Population Synthesis (FSPS) code \citep{con09,con10,for14} with the Padova isochrones \citep{mar07,mar08} and a Chabrier IMF \citep{cha03}. We use the code Prospector \citep{joh21b} for the SED fitting and follow the setup in \citet{lej19b}. We summarize the salient points of the setup here.

We model the SFH with a nonparametric approach. The SFH is divided into 7 time bins, within each bin the SFR is constant. For the SFH priors, we adopt the continuity prior described in \citet{lej19b}, which is tuned to allow similar transitions in SFR to those in hydrodynamical simulations. We adopt the \citet{cal00} dust attenuation curve with extra freedom to alter the overall slope. The optical depths of the diffuse dust and the birth-cloud dust are allowed to differ. For further details on the priors and the model, we refer readers to \citet{lej19b}.

The stellar masses derived from SED fitting are subject to a systematic uncertainty up to $\sim0.2$~dex \citep{lej19b}. The comparison of galaxies in bins of stellar masses in Section~\ref{sec:result} thus carries this systematic uncertainty. We have cross-checked the stellar masses of the same set of galaxies derived from different codes \citep[Prospector, BAGPIPES, MAGPHYS,][]{car18,lej17,dac08} with various settings including SED templates, dust laws, and photometric filters and find scatters of $\sim0.1$~dex with a small systematic offset ($<0.1$~dex) among different sets of measurements (Appendix~\ref{sec:app}). 

To measure the absorption line indices, we first model the emission lines using the Penalized Pixel-Fitting (pPXF) Python routine \citep{cap04,cap17} and subtract these model fits from the observed spectra \citep[see][]{bez18}. Absorption line indices are then measured from the emission-subtracted spectra. 
To estimate the uncertainties on the line indices, we first measure the formal uncertainties based on the noise spectra. However, the noise spectra may not fully capture the full error due to, for instance, flux calibration uncertainties and template mismatches. To estimate the total error budget, for each index, we compare the variances in the differences between galaxies that are observed twice by the LEGA-C survey and have duplicate spectra with $S/N>5$. The scatter between the duplicate measurements is compared to the error estimated from the formal noise spectra. We find that the uncertainties of \dn\ and \ewhd\ from the noise spectra are underestimated by 1.3 and 2.0 times, respectively. We estimate the true typical uncertainties of our \dn\ and \ewhd\ measurements are $\sim0.03$ and $\sim0.55$\AA, respectively.

\subsection{Simulation: IllustrisTNG}
\label{sec:data_tng}

\begin{figure*}
	\includegraphics[width=0.95\textwidth]{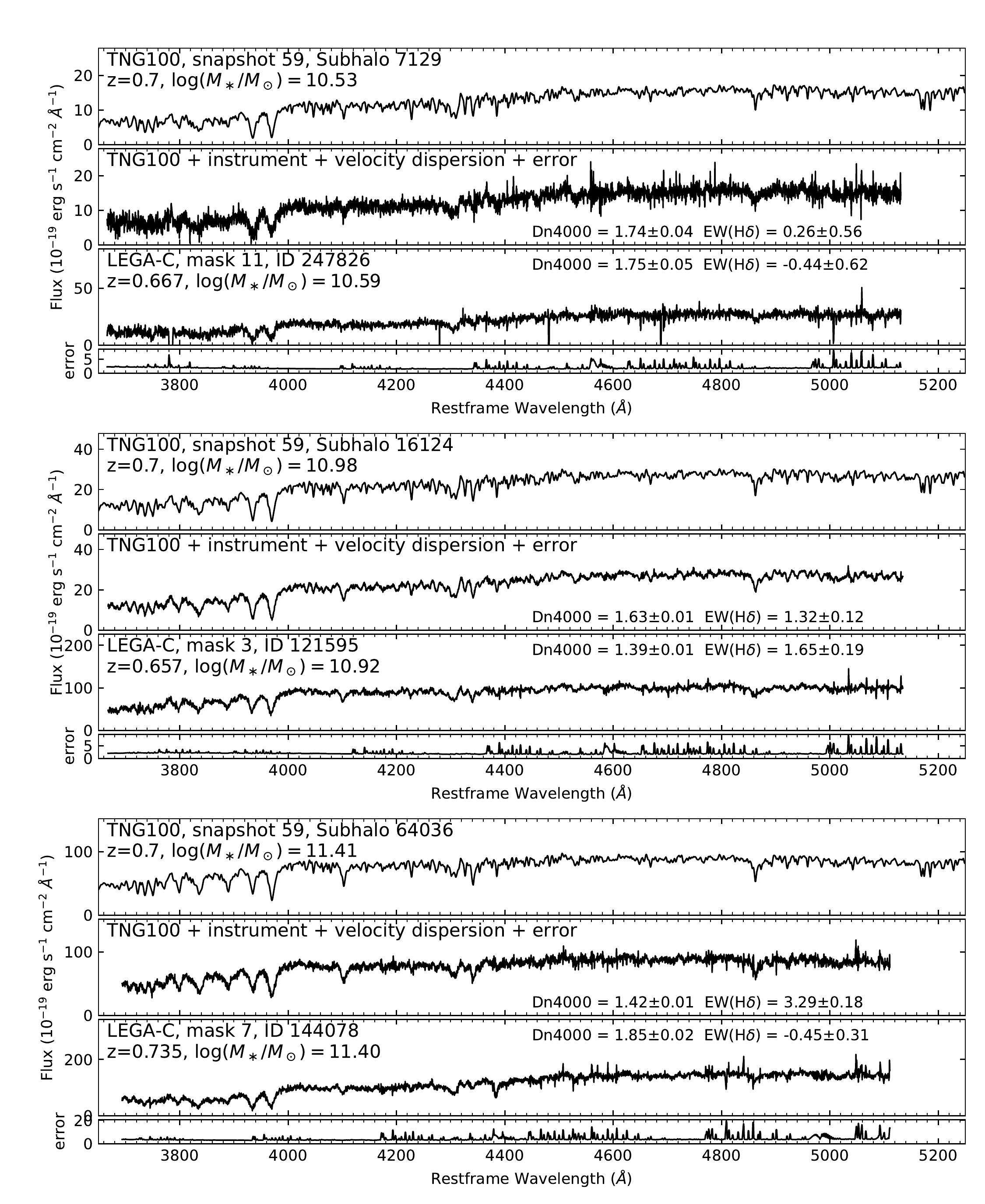}
	\caption{Three examples of TNG100 mock spectra with observational effects matching to the LEGA-C observations. The first row is the original, noise-free mock spectra of a simulated galaxy. We then smooth the spectrum and add noise to produce the spectra of the TNG100 galaxy as if it were observed by the LEGA-C survey (the second row). The width of the smoothing kernel and the noise are based on the stellar velocity dispersion and the noise spectrum of a randomly selected LEGA-C galaxy with similar stellar mass and redshift (see Section~\ref{sec:data_tng}). The S/N ratio is set to be the same as the selected LEGA-C galaxy. The third and the fourth rows are the emission-subtracted spectrum and the uncertainty of a selected LEGA-C galaxy. \label{fig:mock_all}}
\end{figure*}
	
The IllustrisTNG project is a series of cosmological magneto-hydrodynamical simulations incorporating a comprehensive model for galaxy-formation physics \citep{mari18,nai18,nel18,pil18,spr18}. The details of the physical processes and implementation are described in \citet{wei17} and \citet{pil18}. We use the fiducial TNG100 simulation, which has a $\sim100$ comoving Mpc box and a baryon mass resolution of $1.4\times10^6 M_\odot$. 

We generate mock spectra of galaxies (subhalos) in TNG100 using the methodology of \citet{nel18}. Namely, we use the FSPS code with the Padova isochrones \citep{mar07,mar08}, MILES stellar library \citep{san06,fal11}, and a Chabrier IMF \citep{cha03}. We have also repeated the analysis with a Salpeter IMF \citep{sal55} and the conclusions of this paper are unchanged. Each stellar particle is modeled as a simple stellar population (SSP), such that all star formed at the same time with the same metallicity as recorded in the simulation. 

We use the foreground gas cells within 30 physical kpc of simulated galaxies to model the dust attenuation, In brief, for each stellar particle, we use gas cells in the line-of-sight to calculate the column density of the foreground neutral hydrogen and its mean gas metallicity from the metallicity of each foreground gas cell, weighted by the netrual hydrogen mass. We estimate the amount of dust using a metallicity and redshift-dependent dust-to-gas ratio \citep[see][]{mck16}. The attenuation of that stellar particle is then calculated following the prescription of \citet{cal94}. Since we only investigate absorption line indices from stellar light, we do not add line emission to the mock spectra. Summing the attenuated light from all constituent stellar particles makes the mock galaxy spectrum.

To mimic the observations, we first choose a random viewing angle and project the light from each stellar particle over a 2D Gaussian with FWHM of 7.5~kpc on the plane perpendicular to the line of sight, corresponding to the typical seeing of the observations ($\sim1\arcsec$ at $z\sim0.8$). We then apply a rectangular aperture of $7.5\ \mbox{kpc} \times 30\ \mbox{kpc}$ centered at the gravitational center of each subhalo. The apertures are randomly oriented relative to the major axis of the galaxy, mimicking the LEGA-C slit orientations. We sum all of the light within this aperture as the noise-free spectrum (see Fig.~\ref{fig:mock_all}).

Next, we match the observed spectral resolution by convolving each mock spectrum with a Gaussian kernel with a standard deviation (in unit of km~s$^{-1}$) of $\sigma = \sqrt{ \sigma_{gal}^2 + \sigma_{inst}^2 - \sigma_{model}^2 }$. We set the instrument resolution $\sigma_{inst} = 40$~km~s$^{-1}$ and the resolution of the spectral template $\sigma_{model} = 70$~km~s$^{-1}$ \citep{fal11}. The velocity dispersion of the galaxy, $\sigma_{gal}$, is taken from a random LEGA-C galaxy with similar stellar mass ($|\Delta M_\ast| < 0.1$~dex) and similar redshift ($|\Delta z| \leq 0.05$).

Finally, we add realistic noise to the mock spectra. The noise at each wavelength pixel $i$ is drawn from a Gaussian distribution whose standard deviation is $f_{i,mock} / (S/N)_{i,obs}$, where $f_{i,mock}$ is the flux of the mock spectrum and $(S/N)_{i,obs}$ is the S/N of the mass-redshift-matched LEGA-C spectrum at pixel $i$. Therefore, the mock spectrum has the same S/N as the spectrum of the matched LEGA-C galaxy at every wavelength. The absorption line indices are measured from the mock spectrum as with the real LEGA-C spectra.

We calculate the stellar masses of TNG100 galaxies by summing up the masses of stellar particles within 30 physical kpc in a 3D spherical aperture. Such an aperture is shown to yield similar stellar masses to a Petrosian aperture \citep{sch15}. Alternatively, the stellar masses could be derived by fitting mock SEDs to mimic observations. Previous works found small systematic offsets (overall $<0.1$~dex) between the stellar masses from SED modeling and directly summing up stellar particles \citep{low20,abd21}, and we use the latter for simplicity.

\subsection{Constructing samples for comparison}
\label{sec:match}

\begin{figure}
	\centering
	\includegraphics[width=0.95\columnwidth]{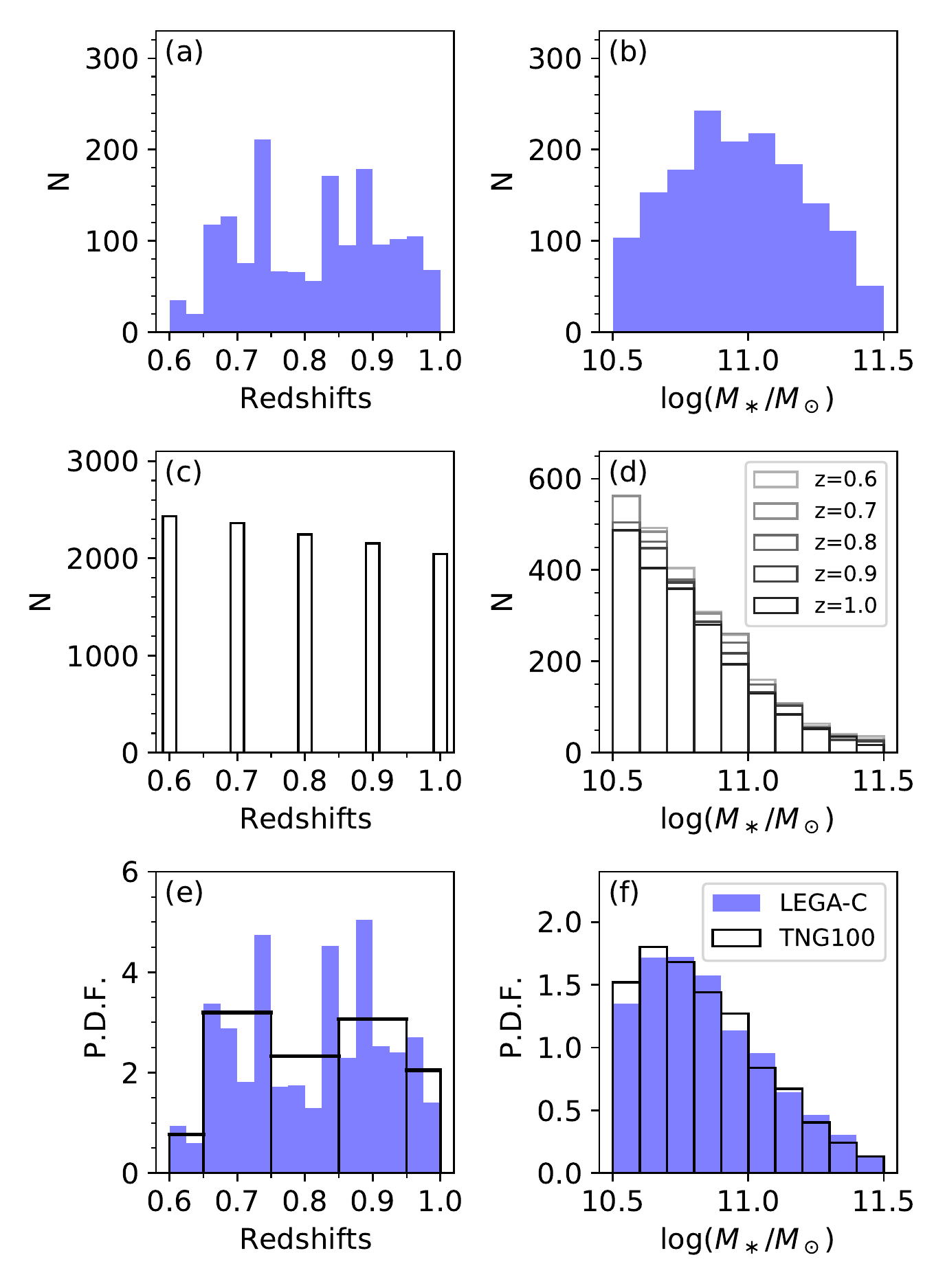}
	\caption{The stellar-mass and redshift distributions of the LEGA-C and the TNG100 sample. Blue histograms are the distributions of the LEGA-C sample before (a,b) and after (e,f) correcting for the completeness of the survey. White histograms in panel (c) and (d) are the total number of galaxies and the stellar mass distribution in each snapshot of the TNG100 simulation. White histograms in Panel (e) and (f) are the redshift and stellar-mass distributions of the TNG100 sample used in this paper. \label{fig:mz_dist}}
\end{figure}

Our aim is to construct samples of galaxies that have similar stellar mass and redshift distributions for the simulation and observation. For the LEGA-C sample, each galaxy is first weighted by the total completeness correction factor $T_{cor}$ \citep{str18} to account for the sampling rate and the standard cosmological volume correction. 
Next, not every observed spectrum covers the wavelength range needed to measure \dn\ and \ewhd. We thus need an additional correction to account for the probability that an observed spectrum can be included for our analysis.

Whether a spectrum covers the necessary wavelengths depends on the redshifts and the position of the galaxy in the slit mask.
The position is independent of galaxy properties, therefore, we only apply a statistical correction for the redshift dependence:
\begin{equation}
I_{cor}(z) = \left \{\begin{array}{ll}
1 ,\quad z \geq 0.8; \\
1 / (3\times z - 1.4),\quad z < 0.8
\end{array}
\right.
\end{equation}
We calculate the fractions of LEGA-C spectra with \dn\ and \ewhd\ at given redshift bins. All spectra at $z\geq0.8$ cover the \dn\ and \ewhd. At $z<0.8$, the probability that \dn\ and \ewhd\ are observed is a linear function of redshift and the reciprocal is the correction factor. 
With the weighting $T_{cor} \times I_{cor}$, we obtain a volumn-complete statistics for the observation. 

For TNG100, we use snapshots 63, 59, 56, 53, and 50, corresponding to $z=0.6, 0.7, 0.8, 0.9$, and $1.0$, respectively. For each LEGA-C galaxy, we randomly draw $5 \times T_{cor} \times I_{cor}$ TNG100 galaxies with similar stellar masses ($|\Delta M_\ast| < 0.1$~dex) from the snapshot with the closest redshift. In this way, we ensure that the LEGA-C and TNG100 samples have similar stellar-mass and redshift distributions. Fig.~\ref{fig:mz_dist} shows the stellar mass and redshift distributions of the LEGA-C and the TNG100 before and after correcting for the completeness of the survey and matching the distributions of stellar masses and redshifts (Fig.~\ref{fig:mz_dist}). We use the samples with survey completeness correction for analysis in this paper. 

The LEGA-C sample contains galaxies in a large-scale structure at $z\simeq0.73$ \citep[see Fig.~\ref{fig:mz_dist}a, and][]{guz07}. In this paper, we do not try to match the local densities of simulated and observed galaxies. We ensure that our results are not severely affected by this high-density region by excluding the member galaxies and repeating our analysis. We find our conclusion in this paper unchanged. We have also repeated the analysis with two subsamples of galaxies with $z<0.8$ and $z>0.8$ and the conclusion is unchanged.

\section{The absorption indices}
\label{sec:result}

\begin{figure*}
	\centering
	\includegraphics[width=0.97\textwidth]{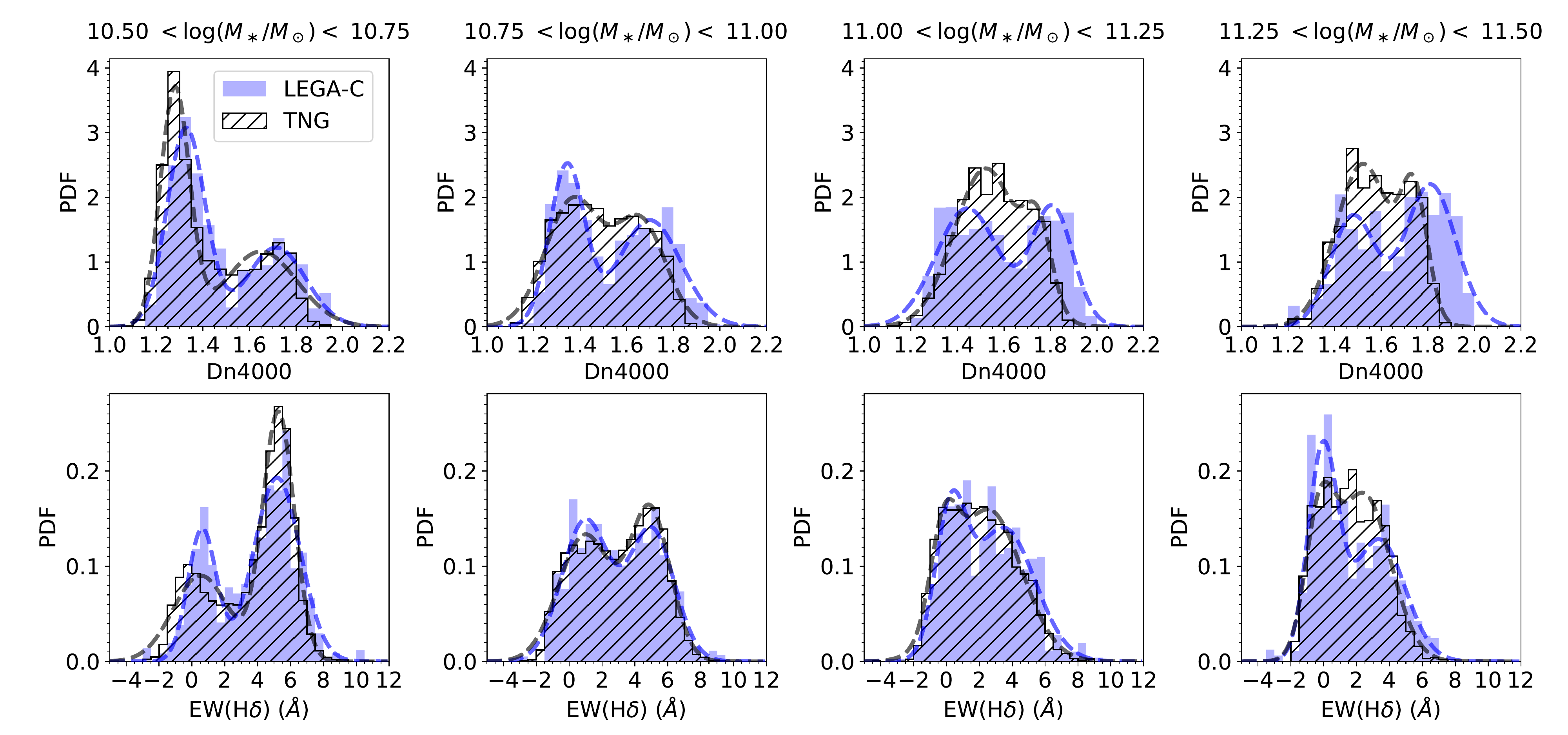}
	\caption{The distributions of \dn\ and \ewhd\ of LEGA-C (blue) and TNG100 (hatched) galaxies in each 0.25~dex stellar mass bin. The dashed curves are best-fit two-Gaussian models for illustrative purposes. A bimodal distribution presents in most of the bins in the LEGA-C sample but only for galaxies with $\log(M_\ast/M_\odot)<11$ in the TNG100 sample.\label{fig:ind_m}}
\end{figure*}

\begin{figure*}
	\centering
	\includegraphics[width=0.97\textwidth]{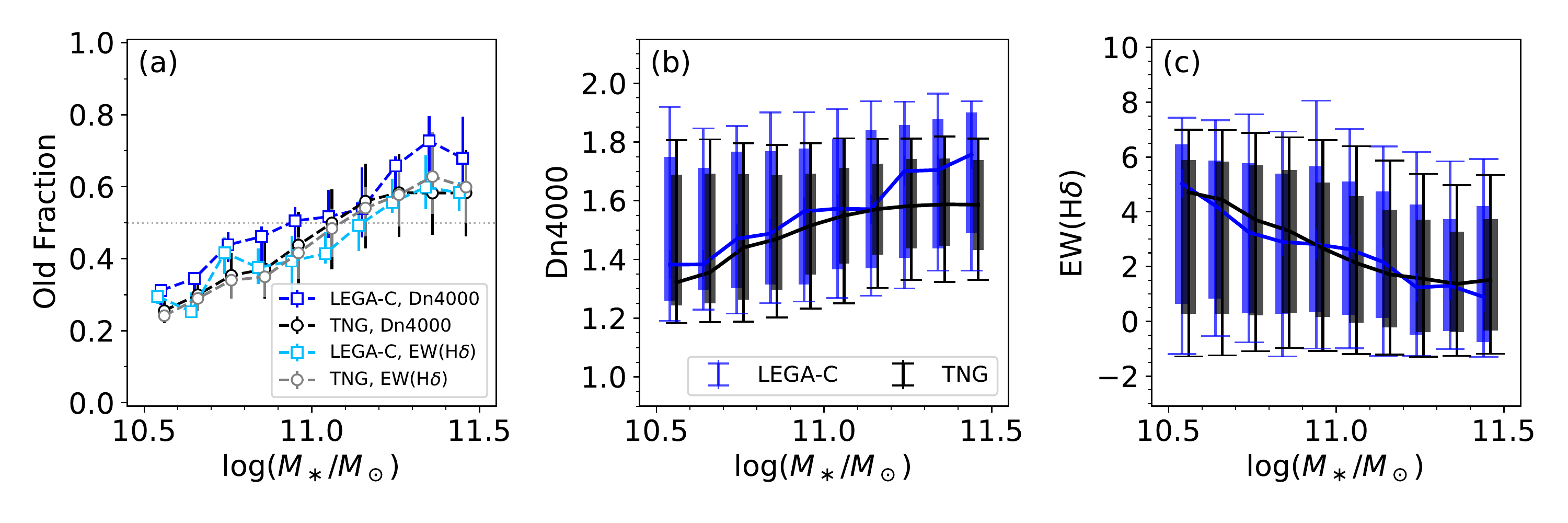}
	\caption{\textit{(a)} The fraction of galaxies with $\mbox{D}_n4000 > 1.55$ or $\mbox{EW(H}\delta\mbox{)} < 2$\AA\ in each 0.1~dex mass bin. The error bars represent the old fractions adopting different boundaries: a 0.05 shift in \dn\ or a 0.5\AA\ shift in \ewhd. At $\log(M_\ast/M_\odot) \gtrsim 11.1$, more than half of galaxies belong to the old population. \textit{(b,c)} The \dn\ and \ewhd\ as a function of stellar masses. The errorbars represent the 2.5, 16, 50, 84, and 97.5 percentiles of the distributions in each bin. The \dn\ of TNG100 galaxies are systematically $\sim0.1$ smaller than LEGA-C galaxies at fixed mass. The \ewhd\ distributions are overall reasonably consistent. \label{fig:index}}
\end{figure*}

The blue histograms in Fig.~\ref{fig:ind_m} are the distributions of \dn\ and \ewhd\ of the LEGA-C sample in each 0.25~dex mass bin. The LEGA-C sample exhibits clear bimodal distributions in most of mass bins. The central troughs fall at $D_n4000 \simeq 1.55$ and $\mbox{EW(H}\delta\mbox{)} \simeq 2$\AA, which separate the `young' and `old' galaxies \citep{kau03a,wu18b}. 

Visually, TNG100 (hatched histograms) reproduces the LEGA-C distributions (blue histograms) reasonably well. This is an important finding, in that it shows that the TNG model returns galaxies at $z\sim1$ whose average age and spread in ages are in reasonably good agreement with observations, also as a function of galaxy stellar mass. However, noticeable differences between the simulation output and the observational results are also in place. In the TNG100 simulation, a bimodality is only present at $\log(M_\ast/M_\odot) < 11$ and vanishes at higher masses. The distributions of TNG100 peak at values of \dn\ smaller than the data at $\log(M_\ast/M_\odot) > 11$, while the \ewhd\ is larger in the most massive bin. 

Taking a heuristic cut at the trough of the LEGA-C population, the \ewhd\ of TNG100 reproduces the observed ratios between the `young' and `old' galaxies for all but the highest stellar masses (Fig.~\ref{fig:index}a). The fraction of old galaxies increases from 30\% in the lowest mass bin to over 50\% at $\log(M_\ast/M_\odot) \simeq 11.1$ and reaches 60\% in the highest mass bin. On the other hand, the old fraction from \dn\ of LEGA-C is systematically $\sim$10\% higher. 

The errorbars in Fig.~\ref{fig:index}b and Fig.~\ref{fig:index}c show the 2.5, 16, 50, 84, and 97.5 percentiles of indices binned every 0.1~dex in stellar mass. As stellar mass increases, TNG100 galaxies have on average larger \dn\ and smaller \ewhd, qualitatively the same as the LEGA-C sample and as expected; more massive galaxies on average consist of older and more metal-rich stars \citep{gal14}. However, \dn\ for the TNG100 sample is systematically $\sim0.1$ smaller. The uncertainties of the percentiles estimated by jackknife resampling are vanishingly small. The offset should reflect a true difference between the mock and observed spectra. 

On the other hand, the \ewhd\ distributions match relatively well. We discuss the details of this agreement further in Section~\ref{sec:sfh_tng}. 

\begin{figure}
	\centering
	\includegraphics[width=0.97\columnwidth]{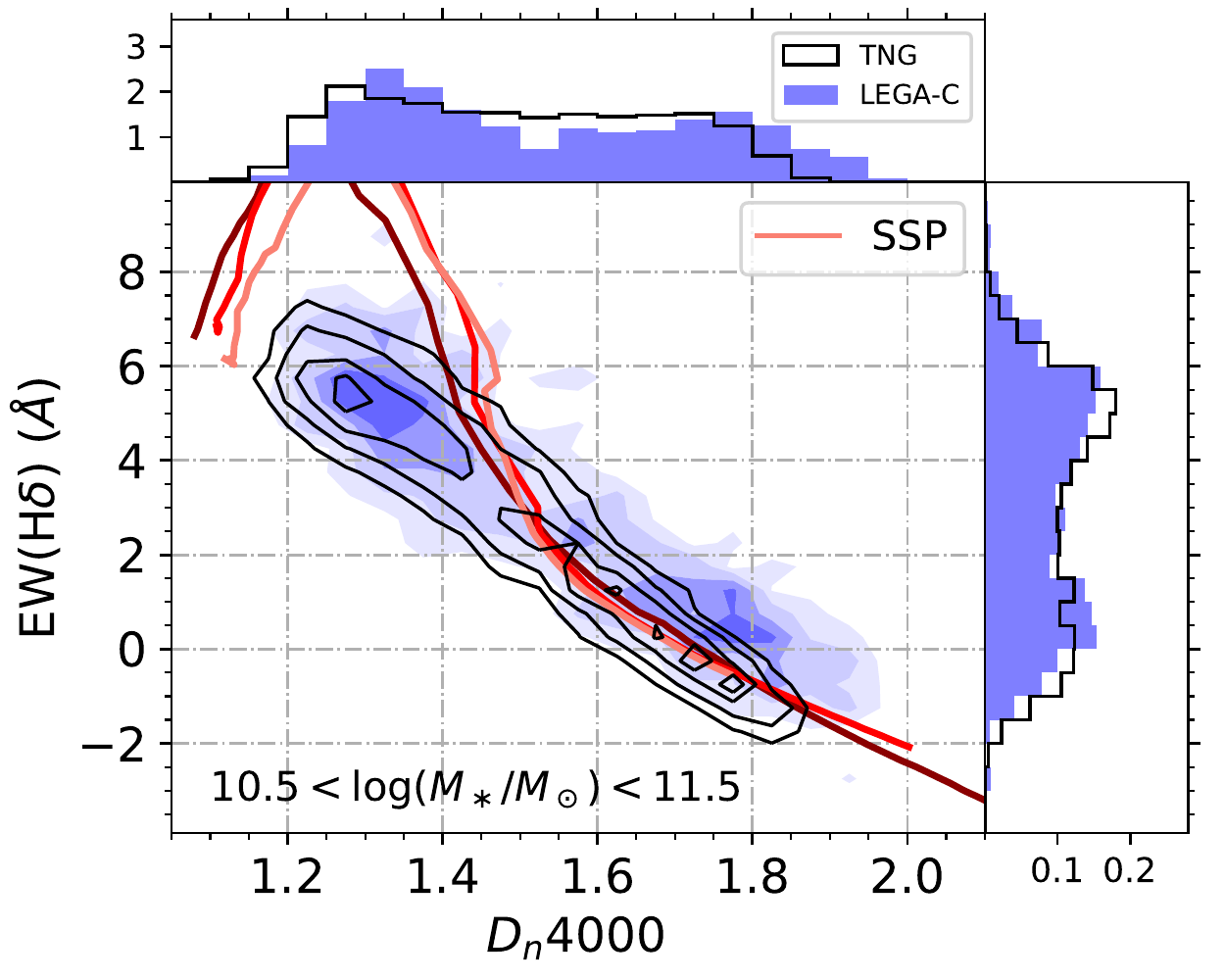}
	\caption{The distributions on the \dn--\ewhd\ plane. The contours represent 0.1, 0.25, 0.5, and 0.75 times of the peak density for the LEGA-C (blue) and TNG100 (black) samples. The upper and right panels are the distributions of \dn\ and \ewhd. Solid lines are dust-free model evolutionary tracks of SSPs generated by FSPS. Lines with dark to light colors represent 2.5, 1.0, and 0.4 times of solar metallicity, respectively. At the large \dn\ end, the LEGA-C sample is located above the TNG100 sample and the SSP model tracks. \label{fig:dnhd}}
\end{figure}

Fig.~\ref{fig:dnhd} shows the distributions of the LEGA-C and the TNG100 samples on the \dn-\ewhd\ plane. There is an overall good agreement, with a small but significant offset between the two samples for old galaxies. For galaxies with large \dn\ and small \ewhd, the TNG100 sample has on average smaller \ewhd\ than the LEGA-C sample at fixed \dn, or equivalently, the TNG100 sample has smaller \dn\ at fixed \ewhd, as can be expected from Fig.~\ref{fig:index}. The locus of old TNG100 galaxies (large \dn\ and small \ewhd) is closer to the dust-free SSP model tracks, while the LEGA-C sample is off. We discuss the possible implications of these mismatches in the next sections. 

\section{Discussion}
\label{sec:dis}
The spectral indices measured from the TNG100 mock spectra reproduce the overall distributions of those of the LEGA-C sample in its qualitative features as well as some of its quantitative aspects (Fig.~\ref{fig:ind_m} and \ref{fig:index}).

The overall agreement implies that TNG100 approximately reproduces star-formation histories of individual galaxy, and their dependence on stellar mass over the first 7 Gyr. On the other hand, the mismatch in \dn\ is statistically significant given the sample size and precision of the measurements, and since the spectra of galaxies evolve slowly at high \dn, small offsets may imply large differences in star-formation histories.

A few potential issues need to be noted before discussing the similarities and differences of stellar populations in the simulated and the observed universes. Foremost, imperfect data reduction, including flux calibration and sky subtraction, could potentially bias or add extra systematics to the index measurements. The data reduction algorithms were notably improved from the second to the final data release, yet the distribution of \dn\ has not changed in a systematic sense. Our extensive, multi-year analysis of the data reduction process leads us to believe that systematic uncertainties do not play a significant role in explaining the \dn\ offset between TNG100 and LEGA-C. That said, this possibility cannot be excluded with 100\% certainty and needs to be kept in mind.

Second, converting from physical quantities of the simulation (ages and metallicities of stellar particles) to spectral observables depends on the adopted models. This systematic uncertainty has to be taken into account. A full exploration of the differences among models is obviously beyond the scope of this paper. In this section, we first point out a few common assumptions in the SPS modeling that are relevant to the results in Sec.~\ref{sec:result} then proceed to the implications on the galaxy formation history from the index measurements. 

\subsection{Conversion between observables and stellar properties}
\label{sec:sps}

\begin{figure}
	\includegraphics[width=0.95\columnwidth]{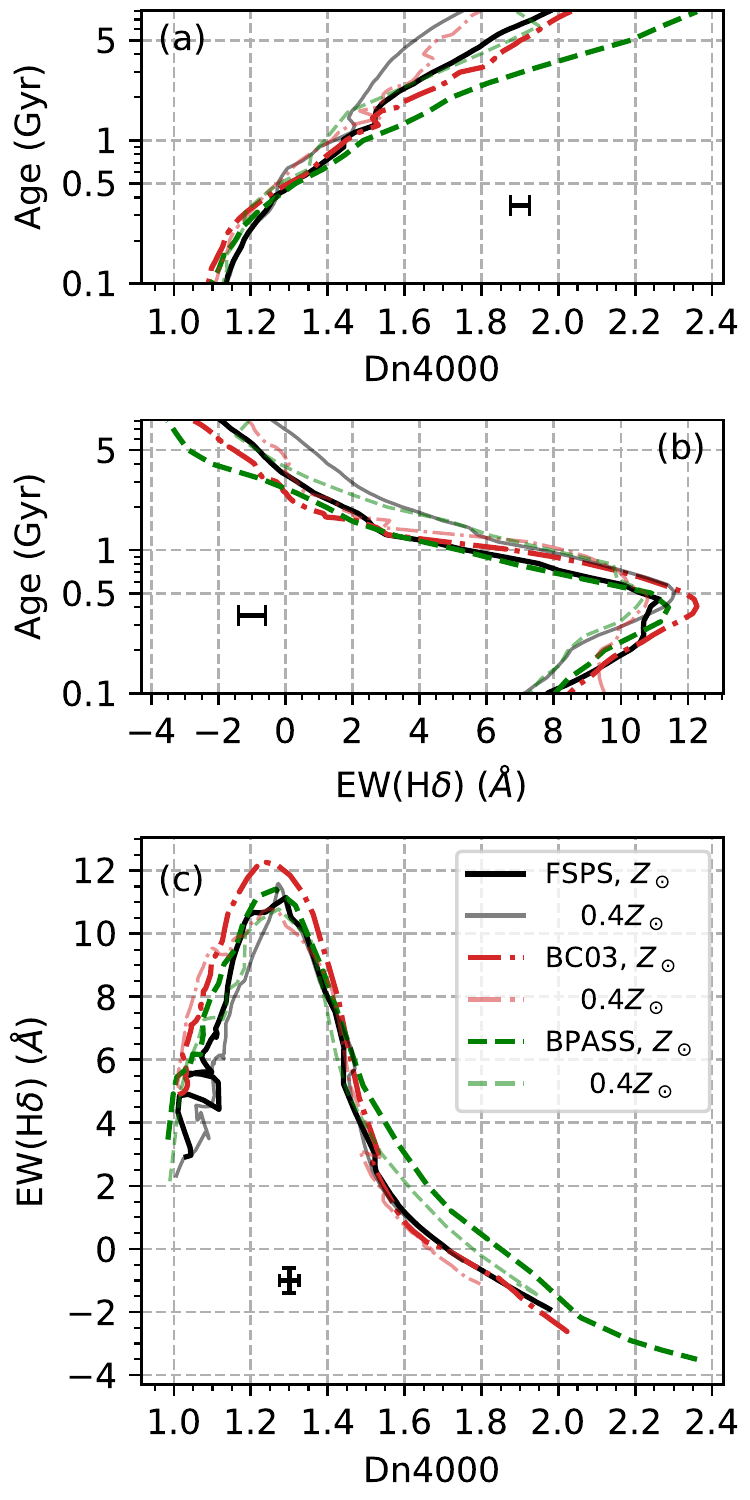}
	\caption{The conversion between spectral indices and stellar ages with different models. Black solid lines are the fiducial FSPS model. Red dotted lines are the BC03 model. Green dashed lines are the BPASS model. Thick lines are with solar metallicity and thin lines are with 0.4 solar metallicity. The errorbars are the typical uncertainties of the LEGA-C sample. With the S/N of the LEGA-C survey, the choice of models can be the dominating source of uncertainty when converting spectral features to stellar ages and metallicities. \label{fig:ssp}}
\end{figure}

As an example, Fig.~\ref{fig:ssp} shows the conversion between the ages and indices of SSPs from commonly used models: FSPS, BC03 (the \citealt{che16} update), and BPASS v2.2 model \citep{eld17,sta18}. All model tracks are with Chabrier IMF, and without dust attenuation. We plot models with solar and 0.4 solar metallicities. For the BPASS model, we plot the models with binary stars and an upper stellar mass cut-off of 300$M_\odot$. These models differ in many aspects, including the stellar evolution after main sequence, stellar spectral templates, and the inclusion of binary stars, etc. The mapping from stellar ages to spectral indices is the combination of all these factors. The indices from the three sets of models differ noticeably at older ages. For instance, a 3-Gyr-old SSP of solar metallicity would have $\mbox{D}_n4000 = 1.67$ with FSPS, $\mbox{D}_n4000 = 1.74$ with BC03, and $\mbox{D}_n4000 = 1.93$ with BPASS. The differences among models are larger than the typical uncertainty of the LEGA-C spectra. There are also noticeable differences in \ewhd. 

Offsets among models also present on the \dn-\ewhd\ plane at all stellar ages (Fig.~\ref{fig:ssp}c). We note that at large \dn, the locus of the dust-free BPASS model is drastically different from the FSPS and BC03 and more consistent with the LEGA-C sample (Fig.~\ref{fig:dnhd}). However, we should not conclude that the BPASS model is overall superior. Firstly, the FSPS models match the distribution of old galaxies in the nearby Universe well on the \dn-\ewhd\ plane \citep{con10}, which means that the BPASS model would be off. Moreover, we will show in Sec.~\ref{sec:sfh} that complex star-formation histories also can shift the distribution on the \dn-\ewhd\ plane.

The abundance ratio is another factor often not taken into account. In this paper and many others, we use scaled solar abundances as our fiducial choice. However, the LEGA-C sample likely spans a wide range of $\alpha$/Fe ratios \citep{str18}, and massive quiescent galaxies may be $\alpha$-enhanced at intermediate redshifts \citep{cho14,ono15,jor17,kri19}. 

\begin{figure}
	\includegraphics[width=0.95\columnwidth]{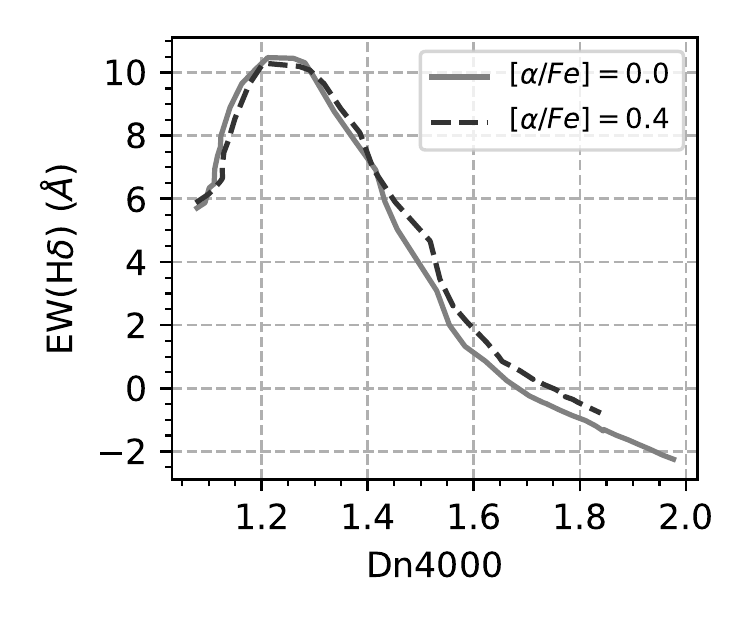}
	\caption{Comparison of evolutionary track of scaled-solar (gray solid line, $[\alpha/Fe]=0.0$) and $\alpha$-enhanced (black dashed line, $[\alpha/Fe] = 0.4$) abundance ratio. Both model tracks are produced with the BasTI isochron $Z/Z_\odot=0.06$, \citet{cha03} IMF, and MILES stellar library. \label{fig:aFe}}
\end{figure}

Fig.~\ref{fig:aFe} shows the evolutionary tracks of SSPs with $[\alpha/Fe] = 0.0$ and $[\alpha/Fe] = 0.4$. Both tracks use the same isochrone \citep[BaSTI,][]{pie04,cor07,pie14}, metallicity ($Z/Z_\odot = 0.06$), IMF \citep{cha03}, and MILES stellar library\footnote{Obtained from \url{http://research.iac.es/proyecto/miles/}}. The $\alpha$-enhanced model is located above the scaled-solar model at large \dn. For old populations, the difference is $\simeq0.05$ in \dn\ or $\simeq0.5$\AA\ in \ewhd. We produce the TNG100 mock spectra with scaled-solar abundance ratios. Qualitatively, the offset between the LEGA-C and the TNG100 samples on the \dn--\ewhd\ may indicate that old quiescent LEGA-C galaxies have $\alpha$-enhanced stellar populations. 

Fig.~\ref{fig:ssp} and Fig.~\ref{fig:aFe} encourage further development in stellar evolution and population synthesis models to improve the accuracy of conversion between observables and physical quantities (ages and metallicities) of stars and galaxies. Full spectral fitting algorithms have improved significantly and now show great successes \citep[e.g.,][]{car18,lej17,rob20,abd21}. When the data quality is improved, e.g., the typical uncertainties of the LEGA-C sample are $\sim0.03$ for \dn\ and $\sim0.4$ for \ewhd, the differences among models become the dominant source of uncertainties, limiting the accuracy of the recovered star-formation histories of galaxies. 

\subsection{The star-formation histories}
\label{sec:sfh}

Let us now assume that systematic errors in the data analysis and the SPS models are not responsible for the \dn\ discrepancies between TNG100 and LEGA-C, and explore potential differences in star-formation history or dust attenuation.

\subsubsection{Simple stellar population}

While galaxies consist of stars of multiple ages and metallicities, and a single absorption feature is clearly not sufficient to infer the underlying age and metallicity distributions, the overall agreement of the \ewhd\ distributions of the TNG100 and the LEGA-C samples is promising, suggesting that the `average' ages and metallicities of TNG100 galaxies should be close to the observed galaxies. 

On the other hand, a naive interpretation for the smaller \dn\ of the TNG100 sample would indicate an overall younger or more metal-poor stellar population. However, if the mismatch is truly from the average ages or metallicity, the \ewhd\ of the TNG100 sample should also be overall larger than the LEGA-C sample, which is not observed. 

Alternatively, a different amount of dust attenuation can in principle explain the shift in \dn\ without affecting the \ewhd significantly. The \dn\ of the LEGA-C sample is on average $\sim0.1$ larger than the TNG100 sample at fixed stellar masses (Fig.~\ref{fig:index}b) and at fixed \ewhd\ (Fig.~\ref{fig:dnhd}). At large \dn, the loci of the TNG100 sample is close to dust-free SSPs (Fig.~\ref{fig:dnhd}), which is consistent with the fact that old quiescent galaxies have low dust attenuation. On the other hand, it requires that the galaxies in the LEGA-C sample have $A_V > 2$~mag to explain the $\sim0.1$ difference in \dn. Such an amount of attenuation is not impossible for star-forming galaxies, but unlikely for quiescent galaxies. A few estimates of the dust attenuation of quiescent galaxies at similar redshifts suggest an $A_V \lesssim 0.5$ \citep{gal14,car19b,bel19}. We constrain the dust attenuation curve and the effect on \dn\ from a joint fit of LEGA-C spectrum and broadband SEDs \citep[following the method in][]{bar20}. We find that only $\sim15\%$ of quiescent galaxies have their \dn\ increased by 0.1 or larger due to dust attenuation. Therefore, the difference in \dn\ is challenging to explain by either the dust attenuation or mismatches in the average age and metallicity. 

\subsubsection{Complex star-formation histories}

\begin{figure}
	\centering
	\includegraphics[width=0.97\columnwidth]{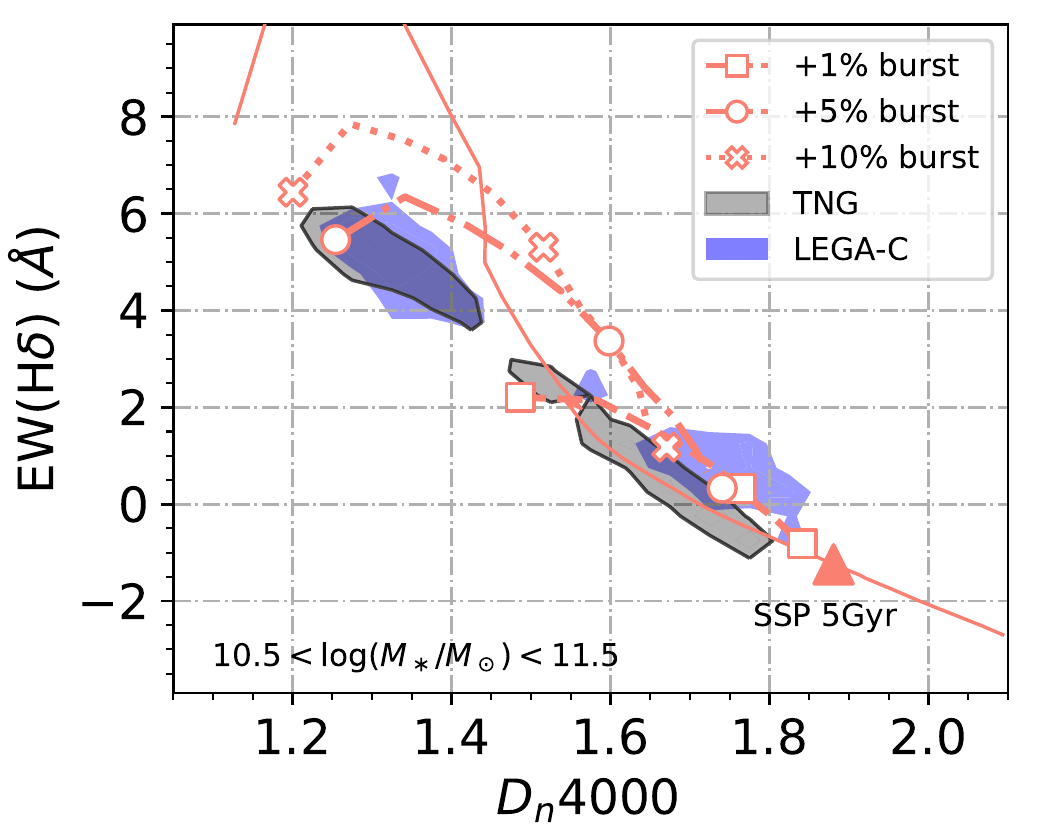}
	\caption{Model tracks with additional bursts on top of an old population (see text). Lines show the evolution from 0.1~Gyr to 1~Gyr after the burst happened. Markers label 0.1, 0.5, and 1.0 Gyr after the burst for the 1\%, 5\%, and 10\% burst models. The solid line is the SSP, equivalent to a 100\% burst event. Whereas the old galaxies in the TNG100 sample (gray) are overall consistent with the SSP model, the LEGA-C galaxies (blue) are more consistent with models with an additional burst that adds a few percent of stellar masses in the last $\sim1$~Gyr. For clarity, only contours of 0.5 times of the peak densities of the simulation and observational galaxy samples are shown -- see Fig.~\ref{fig:dnhd} for the full distributions.  \label{fig:dnhd_burst}}
\end{figure}

Assuming that a simple stellar population cannot explain the \dn-\ewhd\ distribution, we explore the effect of more complex star-formation histories. While the Balmer absorption is mainly contributed from A-type stars, the relatively strong H$\delta$ absorption of the LEGA-C old quiescent galaxies may indicate an excess of relatively young ($\lesssim1$~Gyr) stars. 

For illustration, we plot the evolutionary tracks on the \dn-\ewhd\ plane after new stars formed on top of an existing old population (Fig.~\ref{fig:dnhd_burst}). The models are composed of two SSPs, where the second one is 5~Gyr younger and produces $\leq10\%$ of the stellar mass. All models are generated with solar metallicity and no dust. The extra weak (a few \%) burst effectively produces galaxies with larger \ewhd\ at fixed \dn\ relative to the SSP and match the distribution of the LEGA-C old galaxies. 

\begin{figure*}
	\centering
	\includegraphics[width=0.95\textwidth]{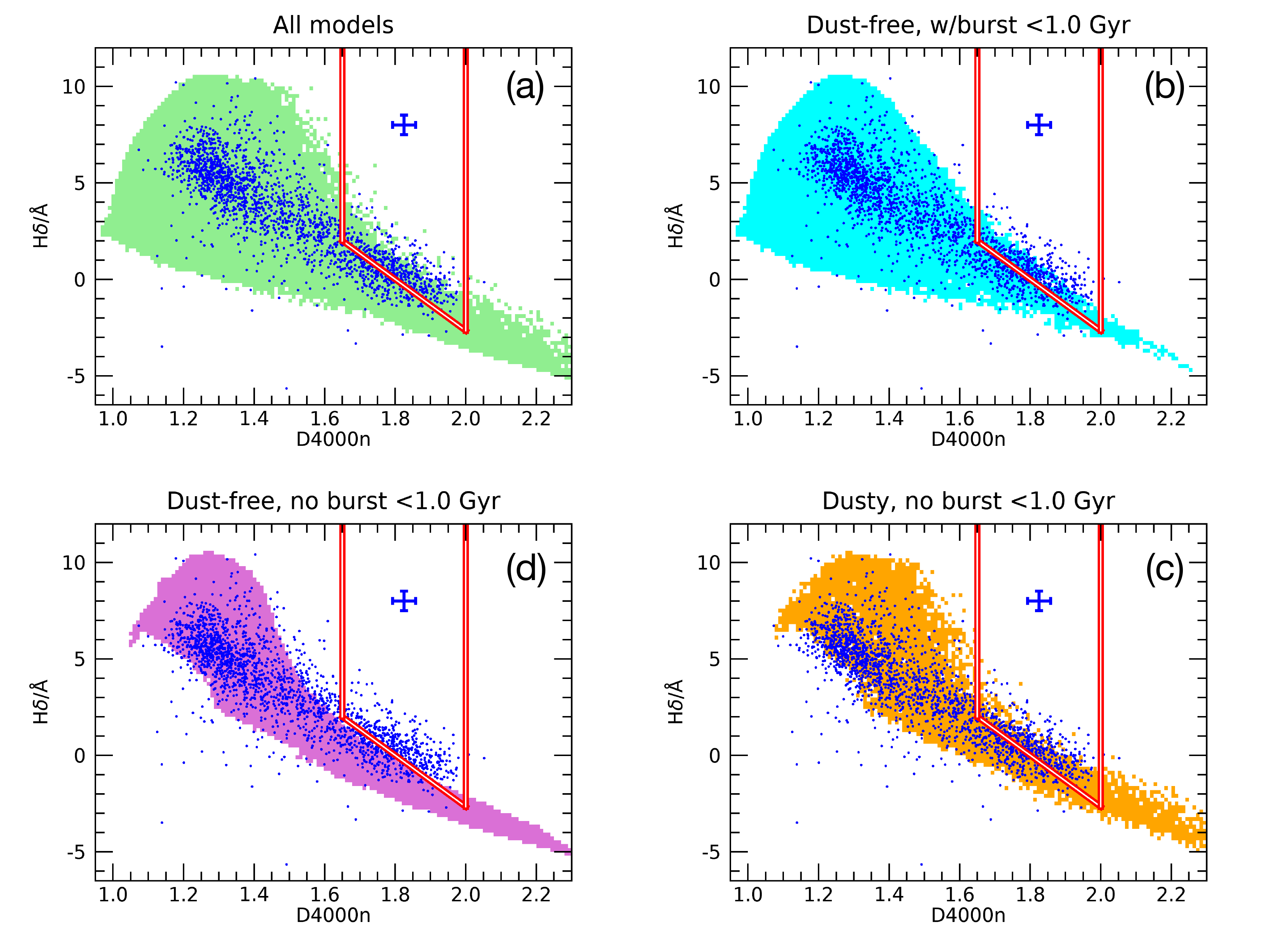}
	\caption{Comparison between the LEGA-C sample and SPS spectral library of \citet{zib17}. Blue dots in each panel are the LEGA-C sample. The errorbars are the typical uncertainties of the LEGA-C spectra. Each model galaxy in the \citet{zib17} library is plotted as a small dot in the background. Galaxies being located outside the colored area cannot, therefore, be matched by any model in the library. \textit{(a):} The entire library. \textit{(b):} Models with low attenuation, $A_V<0.2$. \textit{(c):} Models with $A_V \geq 0.2$ and no burst happened in the past 1~Gyr. \textit{(d):} Models with $A_V<0.2$ and without recent burst. More than half of the LEGA-C galaxies with large \dn\ cannot be matched by models with low attenuation and no recent burst. Either recent bursts or more dust is required. While galaxies with old stellar populations generally have low dust attenuation, the mismatch in panel (d) suggests recent burst events in old galaxies may be common. The slant red line roughly goes through the ridge of the LEGA-C distribution and the upper boundary of the model coverage in panel (d). The star-formation histories of the \citet{zib17} models in the red wedge will be analyzed in Fig.~\ref{fig:zib_burst}.  \label{fig:zib}}
\end{figure*}

Motivated by the illustrative toy models, we then use a more complex spectral library presented in \citet{zib17} to explore the effect of bursts on the evolution on the \dn-\ewhd\ plane. The library consists of galaxies with parametric smooth star-formation history ($SFH(t) \propto \frac{t}{\tau} \exp(-\frac{t^2}{2\tau^2})$) plus multiple bursts, monotonically increasing stellar metallicities, and various amounts of dust attenuation. We refer the readers to \citet{zib17} for a detailed description of the spectral library. Here we limit our analysis to the models whose time elapsed since the beginning of the SFH is $<7.5$~Gyr, corresponding to the age of the Universe at $z=0.7$. 

The library is constructed based on the MILES spectral libraries and \citet{cha03} IMF, which are the same as this paper, but the SPS model is the updated BC03 model \citep{che16}. Since the loci of FSPS and BC03 models on the \dn-\ewhd\ plane are similar (Fig.~\ref{fig:ssp}), we consider the analysis is still informative. 

The \citet{zib17} library allows flexible combinations of formation times, burst strengths, ages, and dust attenuation thus spans a wide area on the \dn--\ewhd\ plane (Fig.~\ref{fig:zib}a). The loci of LEGA-C galaxies are well covered by the \citet{zib17} library. 

The other 3 panels show the coverage of the library 1) excluding galaxies with $A_V > 0.2$~mag (Fig.~\ref{fig:zib}b), 2) excluding bursts within the last 1~Gyr (Fig.~\ref{fig:zib}c), and 3) excluding both high attenuation and recent bursts (Fig.~\ref{fig:zib}d). Models without dust nor recent bursts cannot match most of the LEGA-C galaxies with high \dn\ in the wedge in Fig.~\ref{fig:zib}d. While dust could be a viable solution to match these galaxies, in principle, by shifting the \dn\ to larger values at fixed \ewhd, we do not expect such old galaxies to have significant dust attenuation. Therefore, we conclude that recent bursts are likely the cause of the high \ewhd\ at given \dn. 

\begin{figure*}
	\centering
	\includegraphics[width=0.95\textwidth]{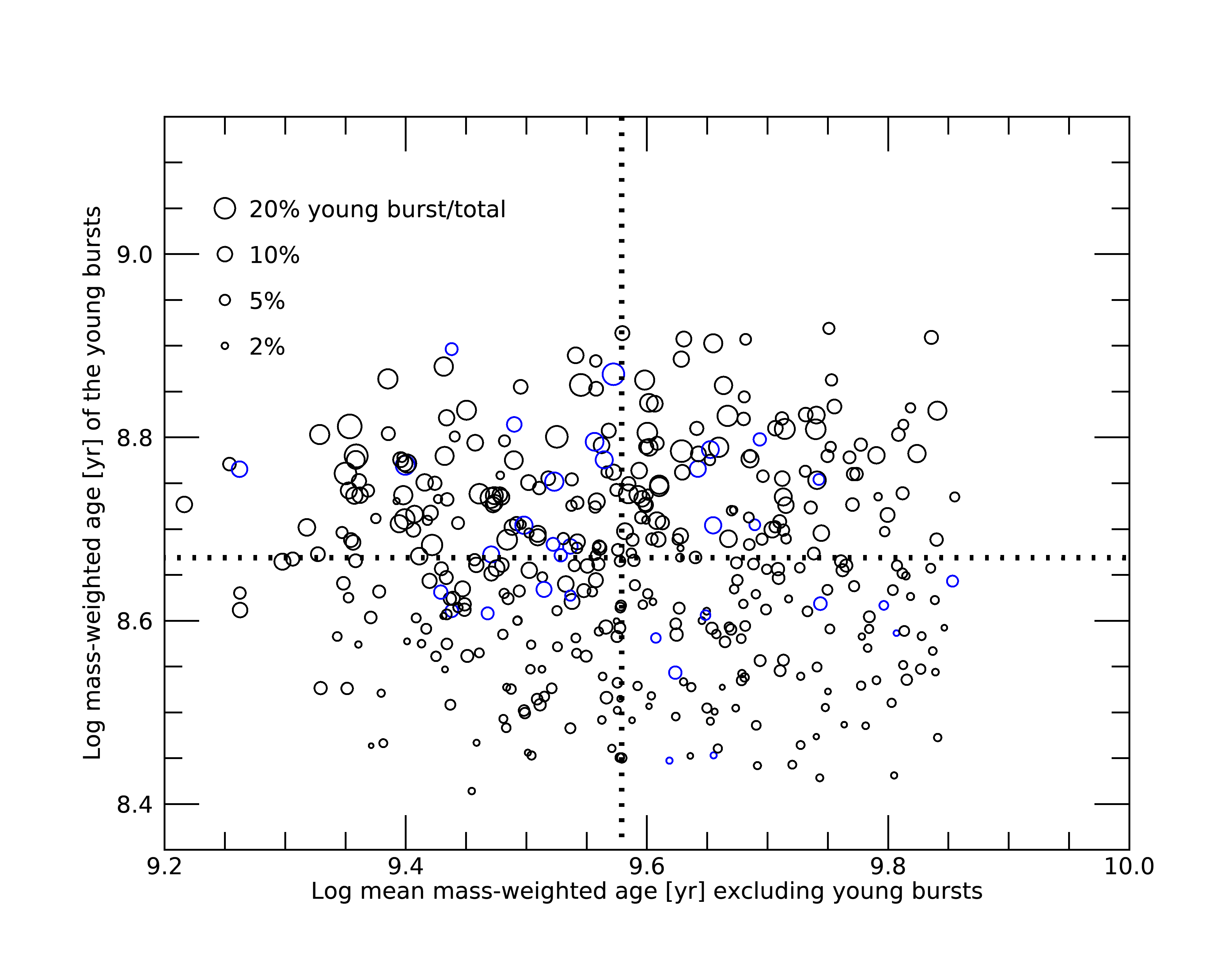}
	\caption{The age and strengths of recent bursts of model galaxies located within the wedge in Fig.~\ref{fig:zib}b. These are the models with low dust attenuation and match the LEGA-C distribution on the \dn--\ewhd\ plane. The vertical axis shows the burst ages and the sizes of the circles represent the fractional masses produced in the bursts relative to the continuous star-formation components (see main text). When there is more than 1 recent burst ($<1$~Gyr), the ages and the strengths are mass-weighted. Models with 1 and 2 recent bursts are in black and blue, respectively. No model has 3 or more recent bursts. The models have a median burst age of $\sim450$~Myr and median burst strength of 3\%. The mass-weighted ages of the model galaxies range from $\sim2$~Gyr to $\sim7$~Gyr. \label{fig:zib_burst}}
\end{figure*}

Assuming negligible dust attenuation ($A_V <0.2$~mag), we investigate quantitatively what ages and strengths of the bursts are needed to reproduce the high \ewhd\ of old galaxies in the LEGA-C sample. We thus select models that are in the wedge marked by the red lines in Fig.~\ref{fig:zib}d and plot in Fig.~\ref{fig:zib_burst} the mass-weighted ages of their bursts in the last 1~Gyr versus the mass-weighted ages of the remaining stars (i.e., the smooth part of the star-formation histories). The symbol sizes represent the percentage of mass in recent bursts relative to the total stellar mass. The star-formation histories of models in the wedge are characterized by an old bulk of stars with mean ages $\gtrsim2$~Gyr up to the age of the Universe (median 3.8~Gyr) with superimposed bursts with an age between 250~Myr and 800~Myr (median 470~Myr), and a mass fraction between a few percents for the youngest bursts up to $\sim20\%$ for the oldest ones (median 3.1\%) \footnote{We have verified that models including bursts older than 1~Gyr do not fall in the wedge. On the other hand, Fig.~\ref{fig:zib_burst} shows that bursts as old as $\sim1$~Gyr can reproduce models in the wedge. Therefore a cut on burst age at 1~Gyr is justified.}.

We note that accurate quantitative statements about the ages and the burst strengths, as well as the ages of the galaxies, may be hindered by the SPS models. Nevertheless, \citet{zib17} libraries and the simple illustrative models in Fig.~\ref{fig:dnhd_burst} tell a consistent story that recent ($\lesssim1$~Gyr), weak (a few percent of total stellar mass) bursts on top of old (a few Gyr) stellar populations can explain the spectral indices of the massive old galaxies in the LEGA-C sample. We have also performed full-spectral fitting on the LEGA-C spectra to constrain the star-formation histories and found $\sim15\%$ of massive quiescent galaxies had at least one rejuvenation event adding $\sim10\%$ of stellar masses \citep{cha18}. Here we further demonstrate that weaker rejuvenation events may be even more common. Although these young stars consist of a small fraction of the total stellar mass budget, their contribution at the rest-frame UV may be detectable \citep[e.g.][]{cos19,sal20,akh21}.

\subsubsection{Formation histories of TNG100 galaxies}
\label{sec:sfh_tng}

Rejuvenation events and their frequency have been investigated in the TNG100 galaxies at $z=0$. In \citet{nel18}, it was found that 10\% of TNG100 red massive ($M_\ast > 10^{11} M_\odot$) galaxies at $z=0$ have undergone a rejuvenation event. Here we investigate the rejuvenation near $z=0.8$. 

We first define `quiescent' galaxies as those whose sSFR drops below $10^{-10.2}\ M_\odot\ yr^{-1}$, averaging over the past 100~Myrs, roughly 0.5~dex below the star-formation main sequence. A rejuvenation event is defined as a period of star-formation activity that brings a quiescent galaxy back to `star-forming' and forms stellar masses above certain thresholds (5.0\% and 1.0\%) of the total stellar mass. In observations, galaxy spectra inform us of the constituent stellar populations but cannot distinguish whether stars form in-situ or ex-situ. Therefore, here we also do not differentiate in- or ex-situ formation for TNG100 galaxies.

\begin{figure*}
	\centering
	\includegraphics[width=0.9\textwidth]{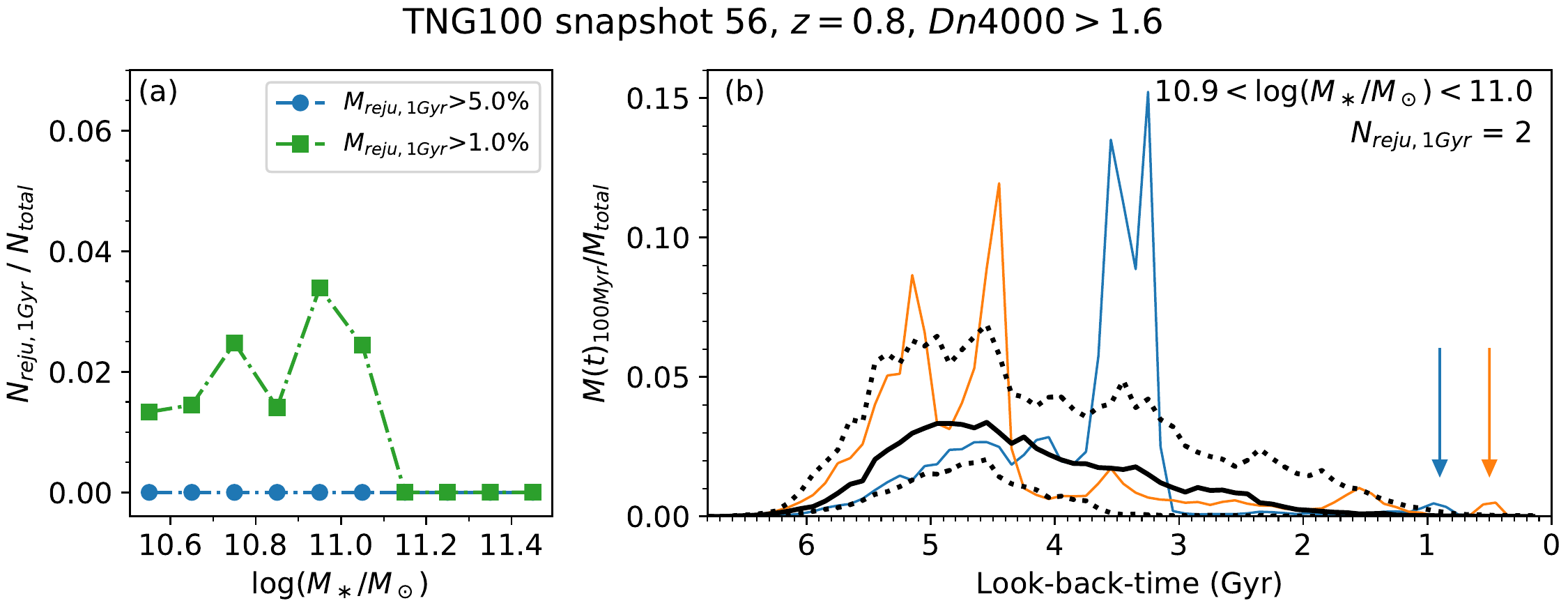}
	\caption{(a) The fraction of old TNG100 galaxies at $z=0.8$ that had a rejuvenation event in the last 1~Gyr as a function of stellar masses. The dash-dotted lines represent the fractions of galaxies with $\mbox{D}_n4000 > 1.6$ that had rejuvenated masses above certain mass thresholds. Only a few percent of old TNG100 galaxies have more than 1\% of their stellar mass formed in a recent rejuvenation event, and none of these events form more than 5\% of the total stellar mass. (b) Age distributions of old TNG100 galaxies at $z=0.8$. The y-axis shows the fractions of stellar mass in each 100~Myr bin compared to the total stellar mass. The thick sold line shows the average age distribution of the constituent stars in TNG100 galaxies with $\mbox{D}_n4000 > 1.6$ and $10.9 < \log(M_\ast/M_\odot) < 11.0$. The dotted lines are the 16th and 84th percentiles. Two galaxies had rejuvenation events that formed $>1\%$ of stellar masses in the last 1~Gyr (thin solid orange and blue lines). The rejuvenation events are indicated by the arrows. \label{fig:reju}}
\end{figure*}

We find that only $\sim2\%$ of old ($\mbox{D}_n4000 > 1.6$) galaxies in the TNG100 sample had rejuvenation events in the past 1~Gyr that formed more than 1\% of the total stellar masses and none of them formed more than 5\% of the total mass (Fig.~\ref{fig:reju}a). For illustration, Fig.~\ref{fig:reju}b shows the age distributions of all old TNG100 galaxies at $z=0.8$ with $10.9 < \log(M_\ast/M_\odot) < 11.0$ and recent rejuvenation events. The dashed lines are the 16th and 84th percentiles of stellar masses in all galaxies at a given look-back time. 

Taken at face value, the rarity of weak recurring star-formation activities may suggest that the feedback prescription implemented in the simulation is too effective: once it kicks in, insufficient further star-formation activity is allowed \citep[also see][for discussion on black hole feedback in TNG100]{ter20}. Alternatively, TNG100 may miss some weak star-formation events that can only take place when the gas is better resolved \citep{don21a,don21b}. If small bursts are related to gas delivered by minor mergers, the gas evolution of satellites could also be involved. The higher-resolution TNG50 simulation \citep{nel19,pil19} will be useful for investigating the effect of the resolution. 

If a small fraction of young ($<1$~Gyr) stars are added to the old TNG100 galaxies, \dn\ will become smaller and \ewhd\ will become larger (see the model tracks in Fig.~\ref{fig:dnhd_burst}), while the total stellar mass is nearly unchanged. The young stellar population will bring the indices of TNG100 galaxies away from the distributions of the LEGA-C sample (see Fig.~\ref{fig:ind_m}). This thought experiment suggests that the bulk of the stars in TNG100 galaxies may be overall younger than those in LEGA-C galaxies, i.e. the quenching mechanism in the simulation may kick in somewhat too late and be too efficient, producing quiescent galaxies with overall younger stars and fewer weak rejuvenation events \citep[see also][]{tacc21} 

\section{Summary and conclusion}
\label{sec:sum}

We present an analysis of \dn\ and \ewhd\ indices, the two most frequently used spectral absorption features in galaxy evolution studies, of galaxies at $0.6 \leq z \leq 1.0$ in the TNG100 simulation. We use the FSPS code to generate the mock spectra and the line-of-sight gas column density to infer the dust attenuation. The sizes of spectroscopic apertures, instrumental resolution, and realistic noise in observations are all taken into account, enabling a detailed comparison of the mock and observed spectra (Fig.~\ref{fig:mock_all}). We compare the simulated indices of TNG100 galaxies with $10.5 \leq \log(M_\ast/M_\odot) \leq 11.5$ against galaxies observed by the LEGA-C survey in the same redshift and stellar mass ranges (Fig.~\ref{fig:mz_dist}). 

TNG100 reproduces the overall distributions of \dn\ and \ewhd\ of the observed galaxy population at fixed stellar mass at $z\sim0.8$ (Figs.~\ref{fig:ind_m} and \ref{fig:dnhd}): this implies that the simulation approximately reproduces the star formation histories of galaxies and their dependence on stellar mass in the first 7 billion years of cosmic evolution. However, noticeable differences are also in place. The simulated indices present strongly bimodal distributions at $ \log(M_\ast/M_\odot) < 11$, while this bimodality becomes weaker than the data at higher masses. The simulated \dn\ is systematically $\sim0.1$ smaller than in LEGA-C at the same stellar mass, while the \ewhd\ distributions differ only at the high-mass end (Fig.~\ref{fig:ind_m} and Fig.~\ref{fig:index}). 

Moreover, LEGA-C galaxies have larger \ewhd\ at fixed \dn\ than the TNG100 galaxies, where this difference is more pronounced for galaxies with old stellar populations ($\mbox{D}_n4000 \gtrsim 1.6$, Fig.~\ref{fig:dnhd}). On the one hand, this tension may be at least partially attributed to the imperfect knowledge in converting from stellar ages and metallicities to spectral indices (e.g. Fig.~\ref{fig:ssp}). Different models treat stellar evolution, stellar spectral templates, and population synthesis differently and result in different mappings from physical quantities to observables. In addition, non-solar abundance ratios are not taken into account.

On the other hand, the differences we find may also suggest that the simulation may need to be refined. The mismatch on the \dn-\ewhd\ plane could be attributed to recent weak rejuvenation events in observed quiescent galaxies. Galaxies with bursts that add a few ($\sim3\%$) of new stars in the recent past ($\sim500$~Myrs) on top of an aged stellar population ($\sim4$~Gyr) can elevate the \ewhd\ at fixed \dn\ and match the locus of the observed galaxy population (Fig.~\ref{fig:dnhd_burst}). Weak rejuvenation events may be common in quiescent galaxies at $z\simeq0.8$. Comparing to the observed galaxies, TNG100 galaxies may turn off star formation relatively late, and once they become quiescent, rejuvenation events are too rare. 

In conclusion, current large observation projects provide high precision measurements of stellar populations for statistical samples of galaxies at large look-back times. Quantitative comparison to state-of-the-art cosmological hydrodynamical simulations is now possible, and allows us to simultaneously test current SPS models, theories of galaxy formation, and the implementation of physical processes in simulations, ultimately leading us towards more accurate galaxy evolution studies. 

\appendix
\section{Stellar masses with different SED fitting methods}
\label{sec:app}

\begin{figure}
    \centering
    \includegraphics[width=0.95\textwidth]{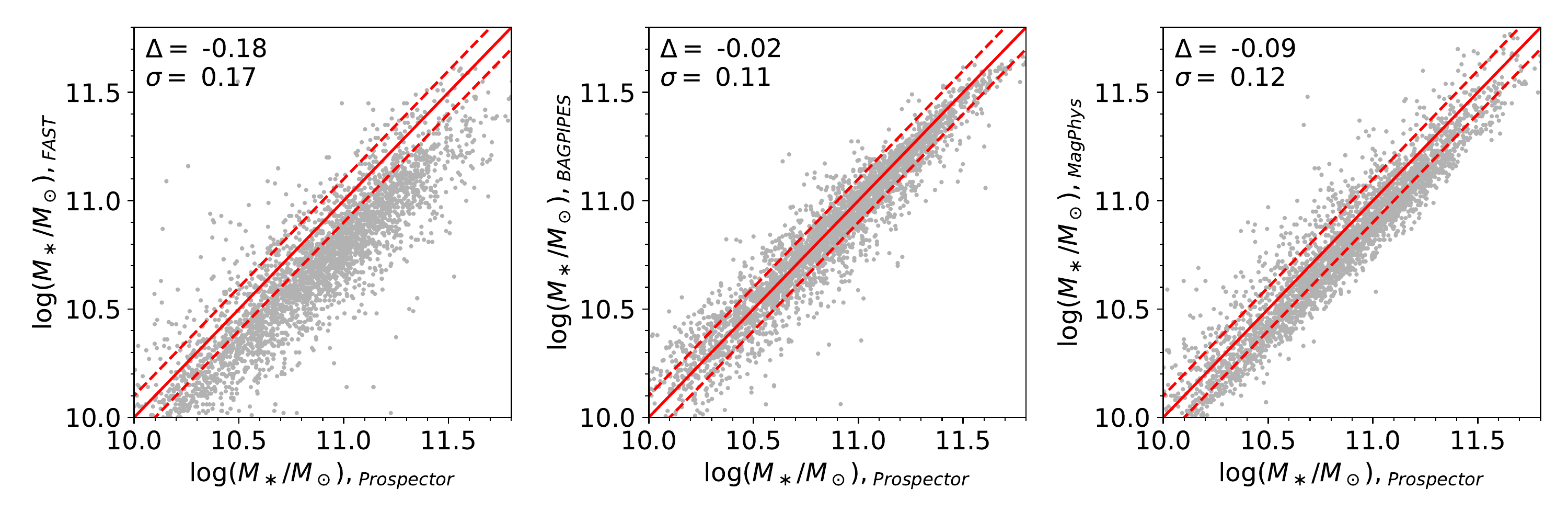}
    \caption{Comparison among stellar masses of LEGA-C galaxies from different SED fitting codes. The stellar masses from FAST (left), BAGPIPES (middel), and MAGPHYS (right) are plotted against the Prospector stellar masses. Red solid lines are the one-to-one relation and red dashed lines show $\pm0.1$~dex offsets. The median offset and scatter are given in the top-left corner. FAST masses are systematically $\sim0.2$~dex smaller than Prospector masses, with $\sim0.2$~dex scatter, as in \citet{lej19b}. On the other hand, the BAGPIPES masses agree with Prospector, and the MAGPHYS masses are systematically $\sim0.1$~dex smaller. The latter two sets have a small scatter of $\sim0.1$~dex in comparison to Prospector.}
    \label{fig:mass}
\end{figure}

In Figure~\ref{fig:ind_m} and Figure~\ref{fig:index}, we show the comparison between the indices of LEGA-C and TNG100 galaxies of similar masses. As the indices depend on stellar masses, the accuracy of the stellar masses from SED fitting may play a role.

In this paper, we use Prospector code for broadband SED fitting and follow the prescription of \citet{lej19b}. Using the 3D-HST catalog \citep{ske14}, \citet{lej19b} showed that their stellar masses are systematically $\sim$0.2~dex larger than the stellar masses derived from the FAST code \citep{kri09}. \citet{lej19b} argued that the exponentially declining SFH models adopted by FAST is oversimplified, which leads to poor representation of SFHs and inaccurate mass-to-light ratios \citep[see also][]{pfo12,car19b,lej19a,low20}.

Besides the fiducial Prospector mass, we also derive the stellar masses of LEGA-C galaxies with various SED fitting codes and settings (Figure~\ref{fig:mass}) using the UltraVista catalog. The first set is the FAST stellar masses, where SED templates are built with \citet{bc03} stellar population libraries, adopting a \citet{cha03} IMF, \citet{cal00} dust law, and exponentially declining SFHs. All available photometry from \textit{GALEX} UV to \textit{Spitzer}/IRAC bands are used in the fits.

The second set is derived with the BAGPIPES code \citep{car18}. While the SPS models, IMF, and the dust law are the same as the FAST setup, there are two major differences. Firstly, we use only $B$ to $J$ broadband filters and the same new photometric zeropoints as for our fiducial Prospector masses (see Section~\ref{sec:data}). We allow the SFR to rise with time by assuming a double-power-law SFH:
\begin{equation}
SFR(t) \propto \left[ \left(\frac{t}{\tau}\right)^\alpha + \left(\frac{t}{\tau}\right)^{-\beta}\right]^{-1},
\end{equation}
where $\alpha$ and $\beta$ are the falling and rising slopes and $\tau$ is related to the time of peak star formation.

The third set is derived with MAGPHYS \citep{dac08}, presented in \citet{gra21}. It uses all available broadbands from $u$ to \textit{Spitzer}/IRAC and \textit{Spitzer}/MIPS 24~$\mu$m, the new photometric zeropoints, \citet{bc03} with a \citet{cha03} IMF, and the two-component dust model from \citet{cf00}. The SFHs are exponentially declining with additional random bursts.

There is a $\sim0.2$~dex offset between the Prospector and FAST masses, similar to \citet{lej19b}, despite that we use fewer broadband filters than \citet{lej19b}. The BAGPIPES and Prospector masses, while based on different SPS models and assumptions of SFHs, agree with each other very well. The MAGPHYS masses is on the other hand slightly smaller.

We find that the 3 sets of stellar masses from more complex SFHs (Prospector, BAGPIPES, MAGPHYS) show a high degree of consistency: $<0.1$~dex offset and $\sim0.1$~dex scatter. On the contrary, the FAST stellar masses, based on exponentially declining SFHs, are systematically smaller by $\sim0.2$~dex and have larger scatter ($\sim0.2$~dex) when compared to the Prospector masses. Deriving physical parameters from broadband SED fittings is a complex, high-dimensional problem. The simple comparison here is not meant to be a thorough analysis but serves as an illustrative case. The overall result is consistent with previous investigations, which in general suggest that complex SFH models improve SED fittings \citep{pfo12,car19b,lej19a,low20}. We thus adopt the Prospector stellar masses, which are based on the same population synthesis model as our mock spectra.

\software{Astropy \citep{astropy13,astropy18}, BAGPIPES \citep{car18}, FSPS \citep{con09,con10}, MAGPHYS \citep{dac08}, pPXF \citep{cap04,cap17}, pyFSPS \citep{joh21a}}

\acknowledgments
We thank the referee for the valuable comments. This work is based on observations made with ESO Telescopes at the La Silla Paranal Observatory under programme ID 194-A.2005 and 1100.A-0949 (The LEGA-C Public Spectroscopy Survey). This project has received funding from the European Research Council (ERC) under the European Union’s Horizon 2020 research and innovation program (grant agreement No. 683184). P.F.W. acknowledges the support of the fellowship by the East Asian Core Observatories Association. A.G. acknowledges support by the INAF PRIN-SKA2017 program 1.05.01.88.04.ESKAPE-HI. KEW gratefully acknowledges funding from the Alfred P. Sloan Foundation.

\bibliography{TNG_LEGAC_v4_arXiv_R2}
\end{document}